\documentclass[]{aastex631}
\usepackage{booktabs}
\usepackage{amsmath}

\begin{document}

\title{Tides from the cloud can induce the fast disruption of star clusters and offer an explanation for Gaia strings}

\author{Xiao-Tong Chen}
\affiliation{South-Western Institute for Astronomy Research\\ Yunnan University\\ Chenggong District\\ Kunming 650091, P. R. China}

\author{Guang-Xing Li}
\affiliation{South-Western Institute for Astronomy Research\\ Yunnan University\\ Chenggong District\\ Kunming 650091, P. R. China}
\email{gxli@ynu.edu.cn}






\begin{abstract}
Young stars form in clusters within molecular clouds, but older stars are evenly distributed across the galactic disk, necessitating an explanation for cluster dissolution. We analytically study tidal forces from cold molecular clouds as a key mechanism for accelerated cluster disruption. Cloud tides, caused by the gravitational pull of the parent cloud along the radial direction, arise from the spatial gradient of gravitational acceleration and drive cluster disruption. This mechanism activates after gas expulsion and remains effective until the cloud is disrupted by stellar feedback or the cluster moves away.  Cloud tides act on gas-deprived clusters, causing exponential expansion on a tidal timescale of $t_{\rm tidal,ext} = \sqrt{3/(8\pi G\rho_{\rm mean})}$, where $\rho_{\rm mean}$ is the cloud’s density at the cluster’s location. With a duration of a few Myr, cloud tides can lead to a 10 times increase of the cluster size,  producing bar-like elongated stellar aggregations resembling Gaia strings. These results establish cloud tides as a potentially important mechanism for star cluster disruption.

\end{abstract}

\keywords{Cloud tide --- Dynamics --- Cluster evolution}


\section{Introduction}


Stars distribute hierarchically—clustered in galaxies, spiral arms, and nebulae \citep{Kruijssen_2012,Elmegreen_2018}. Young stars emerge primarily at the high-density clump in dense molecular clouds.  How star clusters dissolve from this clustered distribution at their birth to a relatively uniform distribution remains an open question. Observations indicate that most young stars are found in clusters; however, the majority of ``normal" stars in the Milky Way are not clustered. Several mechanisms have been proposed to explain the dissolution of the star clusters:

\begin{itemize}
    \item \textbf{Gas expulsion  (\citealt{1980ApJ...235..986H}; \citealt{10.1046/j.1365-8711.2003.06076.x})}: The injection of velocity due to the change of gravitational potential during gas removal.  
    Only a fraction of the gas in a clump can be converted to a star cluster, with the rest of the gas being expelled. This leads to a shallower gravitational potential and a relatively slow expansion velocity of 1-5 kilometers per second. \cite{Baumgardt_2007} conducted detailed N-body simulations and demonstrate that a cluster's survival of gas expulsion is primarily controlled by the star formation efficiency and the speed of gas removal.
    \item \textbf{Scattering by molecular clouds.} When a star cluster passes through a cloud, the rugged gravitational potential of the cloud can inject different momenta to different stars, leading to larger velocity dispersions \citep{1958ApJ...127...17S,theuns2002numericalsimulationsintergalacticmedium}. The cluster lifetime due to periodic heating from passing clouds is inversely proportional to the volume density of molecular gas and proportional to the density of the cluster, and these encounters can result in waves which decelerate the star cluster \citealt{1972ApJ...176L..51O}. These encounters can also lead to the production of molecular contrails, which are elongated concentrations of cold molecular gas \cite{2021MNRAS.503.4466L}. However, this process is considered slow, with a typical timescale of  1-10 Gyr \cite{2019ARA&A..57..227K}.
    \item \textbf{Galactic tides (Gyr).}
    External processes are integral to the evolution of star clusters, notably tidal disruption. Clusters do not exist in isolation but are influenced by the local tidal field of their host galaxy
    \citep{Portegies_Zwart_2010}. In Baade's \citep{1934BHarO.895....1B} investigation, stellar groups are found to be unstable against Galactic tidal forces if their mass density is lower than 0.1-0.3 $M_{\odot} \, {\rm pc}^{-3}$. As a long-range force, the Galactic tide can stretch clusters over timescales of a few gigayears, which is comparable to the scattering time with clouds. 
\end{itemize}

According to the literature, after gas expulsion, the dissolution of the star cluster is a slow process occurring on the timescale of around Gyr. However, the Gaia mission identified extraordinary star clusters known as ``Gaia strings" \citep{2019AJ....158..122K}, which are young stellar associations characterized by their elongated,
string-like shapes. The Gaia strings have velocity dispersions too large to be
explained by the cloud turbulence, and this velocity discrepancy (Fig.
\ref{fig:gaia}) is too large to be explained by the intrinsic velocity
dispersion of the gas.
The stars in the strings are young (around 100 Myr), this timescale is too short
for effects such as scattering and Galactic tide to be effective. Some
additional mechanisms are needed to explain the sizes and velocities of these objects.



We investigate the effect of tidal forces from the parental molecular cloud on the disruption of star clusters. The molecular clouds have a hierarchical density distribution, with star clusters forming at parsec-scale density peaks. Several star clusters may form in one cloud. When a star cluster has formed, gravity from the rest of the cloud should affect the evolution of the star cluster, potentially leading to dissolution. Tidal forces have long been considered a potential mechanism leading to the disruption of star clusters; however, previous studies primarily focused on the Galactic gravitational field as the source of these tides, resulting in very long disruption timescales ($\approx$ Gyr), suggesting a slow influence.

\begin{figure}
    \centering
    \includegraphics[width=0.85\linewidth]{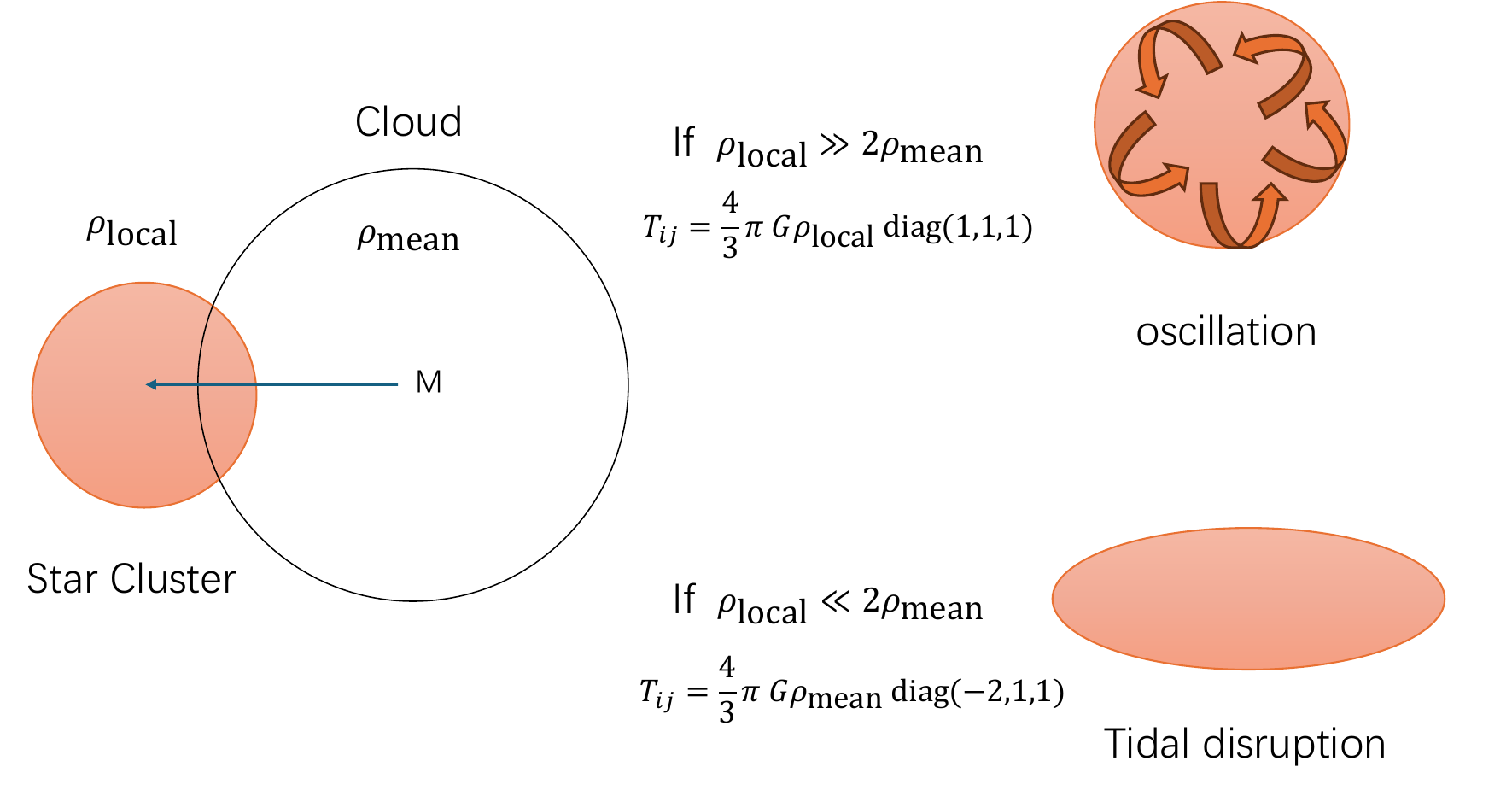}
    \caption{An illustration of cloud tide. The sketch shows a cluster with density $\rho_{\rm local}$ different evolution under cloud tide. Befor gas expulsion, the tidal tensor is compressive. After gas expulsion, cluster could be disrupted by its parent cloud.}
    \label{fig:enter-label}
\end{figure}

We propose that tidal force from the parent cloud can lead to a fast disruption of the star cluster. 
In Figure \ref{fig:enter-label} we illustrate the mechanism of cloud tide. Here, a star cluster is under the influence of the tide from a nearby cloud. We analytically study the evolution of a star cluster under the tide from a nearby gas cloud, and compare our results with observations.


\section{Tidally-driven cluster expansion}



\subsection{Mechanism}
We study the effect of tidal force on a star cluster located at a distance $r$ from a cloud of mass $m$, and as the tidal force could have an acceleration effect, the cluster could be stretched by $x$. 
Tidal force is related to the spatial gradient of the gravitational acceleration. Assuming that the potential $\Phi$ at any point is related to the local density $\rho$ by Poisson  equation,
\begin{equation}
    \nabla^{2}\Phi = 4\pi G \rho \;.
\end{equation}

We quantified the velocity and length scales of the cluster, where the tidal tensor in the spherical coordinate 
\begin{equation}
 T_{ij}= {\rm diag} (\lambda_r, \lambda_\theta, \lambda_\phi ) \; ,
\end{equation}

Assuming that the location of the cluster still with some gas has a density of $\rho_{\rm local}$, and molecular cloud provide an external tides with a density of $\rho_{\rm mean}$, the tidal tensor should be
\[
T_{ij} = \frac{4}{3} \pi G \rho_{\rm mean}
\begin{bmatrix}
-2 & 0 & 0 \\
0 & 1 & 0 \\
0 & 0 & 1
\end{bmatrix}
+\frac{4}{3}\pi G \rho_{\rm local}
\begin{bmatrix}
1 & 0 & 0 \\
0 & 1 & 0 \\
0 & 0 & 1
\end{bmatrix}
\;,
\]
Before gas expulsion, $T_{ij}$ is compressive and dominated by $4/3(\pi G \rho_{\rm local}) \rm{diag}(1,1,1)$, the cluster oscillate under its self-gravity. When gas completely expelled, we find $T_{rr}$ can increase the disruption of cluster.

Thus,
\begin{equation}
    a_{i} = -\frac{\partial \phi}{\partial x_{i}} \;,
\end{equation}
\begin{equation}
    T_{ij} = -\frac{\partial^2 \phi}{\partial x_{i} \partial x_{j}} = \frac{\partial a_{i}}{\partial x_{j}} \;.
\end{equation}


The component $T_{rr}$ represents the tidal tensor along the radial direction and calculate as $ \frac{2 GM}{r^{3}}$ . To simply calculate $T_{rr}$, we assume an average density ${\rho} =  \frac{m}{\frac{4}{3}\pi r^3}$, which is the characteristic density of the cloud measured at the location of the cloud, where $T_{rr} = \frac{8\pi}{3}G {\rho}$. 
\subsection{Radial expansion} \label{sec:radial}
Assuming $v$ is the expansion velocity, the evolution of the cluster is described by  
\begin{equation}
    \frac{dx}{dt} = v \;,
\end{equation}
\begin{equation}
    \frac{dv}{dt} = \Delta a \;,
\end{equation}
\begin{equation}
    \Delta a = T_{ rr} \, \Delta x=\frac{GM}{x^{2}}=\frac{8\pi}{3}G {\rho}x \;.
\end{equation}




\section{Cluster Evolution Under Tidal Forces}

We define the tidal timescale of external, as
\begin{equation}
t_{\rm tidal,ext} = \sqrt{\frac{3}{8\pi G \rho}}\;,
\end{equation}
for a cluster with initial velocity $v_0$ and length variation $x_0$. The solution becomes
\begin{equation}
x = \frac{1}{2} \left( x_0 + v_0 t_{\rm tidal,ext} \right) e^{\frac{t}{t_{\rm tidal,ext}}} + \frac{1}{2} \left( x_0 - v_0 t_{\rm tidal,ext} \right) e^{-\frac{t}{t_{\rm tidal,ext}}}\;,
\end{equation}
\begin{equation}
v = \frac{1}{2} \left( \frac{x_0}{t_{\rm tidal,ext}} + v_0 \right) e^{\frac{t}{t_{\rm tidal,ext}}} - \frac{1}{2} \left( \frac{x_0}{t_{\rm tidal,ext}} - v_0 \right) e^{-\frac{t}{t_{\rm tidal,ext}}}\;.
\end{equation}

\subsection{Tangential Oscillation}

Along the tangential direction, the evolution is described by
\begin{equation}
\ddot{x} = -\frac{4\pi}{3} G {\rho} x\;.
\end{equation}

The solutions should be
\begin{equation}
x = x_0 \cos\left( \frac{t}{\sqrt{2}t_{\rm tidal,ext}} \right) + \sqrt{2}v_0 t_{\rm tidal,ext} \sin\left( \frac{t}{\sqrt{2}t_{\rm tidal,ext}} \right)\;,
\end{equation}
\begin{equation}
v = v_0 \cos\left( \frac{t}{\sqrt{2}t_{\rm tidal,ext}} \right) - \frac{x_0}{\sqrt{2}t_{\rm tidal,ext}} \sin\left( \frac{t}{\sqrt{2}t_{\rm tidal,ext}} \right)\;.
\end{equation}
In the tangential direction, clusters mainly show oscillatory behavior.

\subsection{Initial Tidal Virialization}

\begin{figure}
    \centering
    \includegraphics[width=0.85\textwidth]{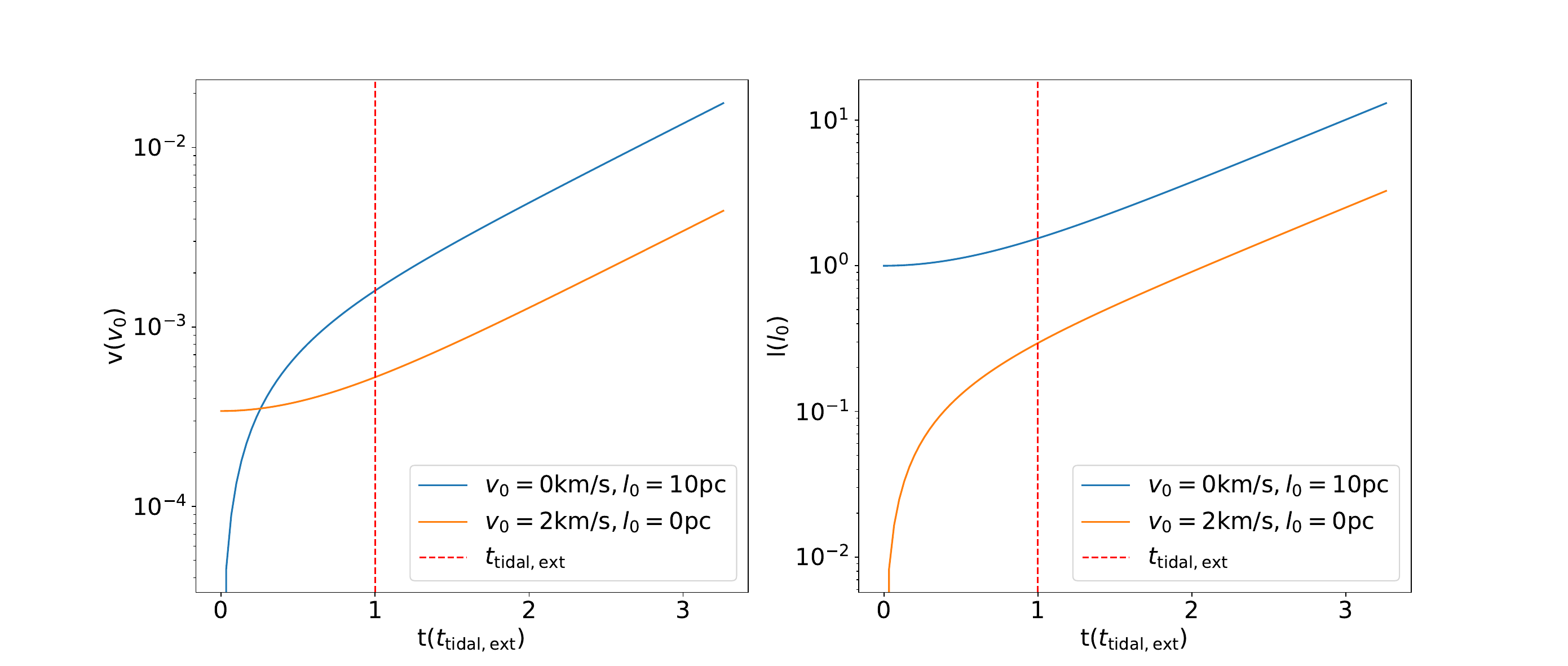}
	\caption{Cluster evolves with size-dominated (blue line) and velocity-dominated (orange line) initial preconditions. With a tidal timescale, the cluster is virialized.} 
	\label{fig:3}
\end{figure}

Figure \ref{fig:3} shows the evolution of a cluster in the vicinity of a molecular cloud with an average density of ${\rho}$.
The effective length of the cluster is defined as
\begin{equation}
l_{\rm eff} = \sqrt{l_0^2 + (v_0 t_{\rm tidal,ext})^2}.
\end{equation}
This represents a stationary length when a star cluster has virialized (e.g., a large initial velocity is converted into size and vice versa) under tidal forces.

This equation has two representative cases:
\begin{itemize}
\item Size dominated: The cluster will acquire velocity gradually through the tide.
\item Velocity dominated: The cluster will reach the size of $l_{\rm eff}$ through free expansion, after which the expansion will be driven by cloud tide.
\end{itemize}
The star cluster will reach this size after a timescale of $t_{\rm tidal,ext}$.



\section{Simulation and observation of cloud tide}
\subsection{Complete picture cluster evolution}\label{sec:complete}



\begin{figure}
    \centering
    \includegraphics[width=0.85\textwidth]{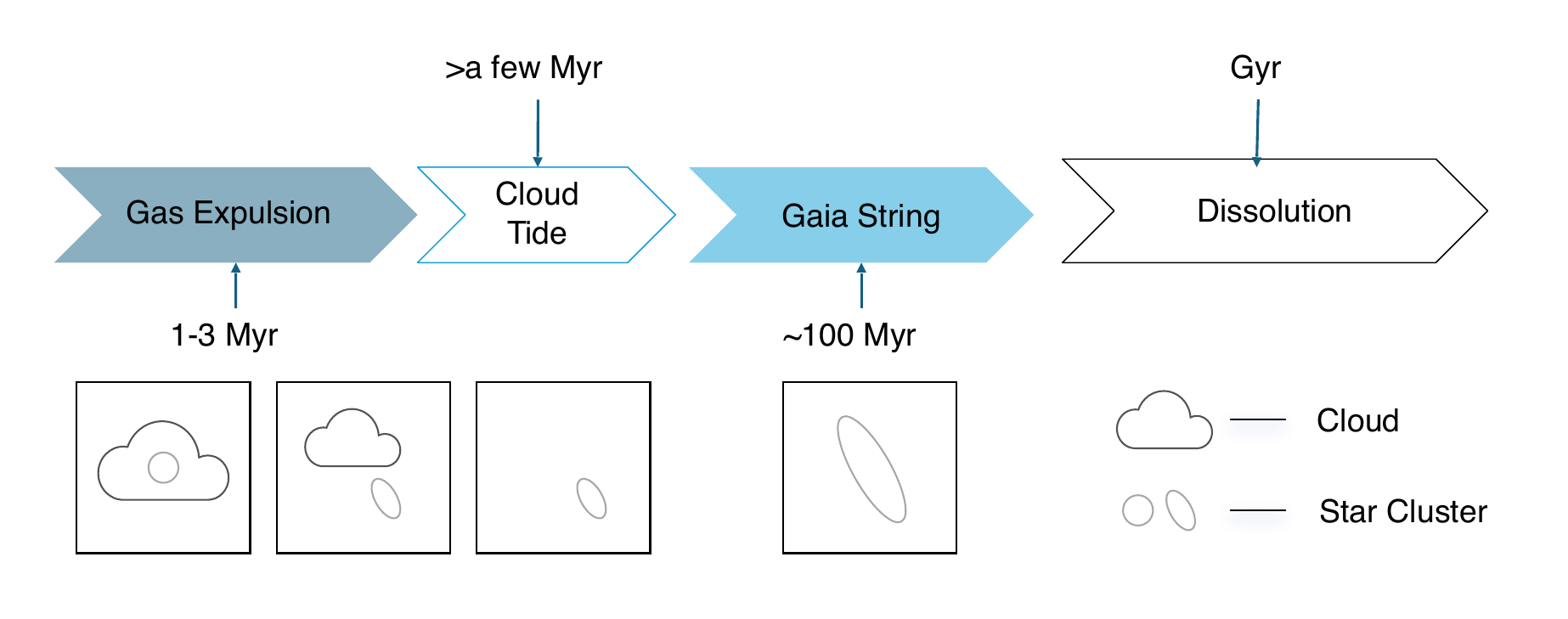}
	\caption{Diagram showing the entire evolutionary process of the cluster. Star clusters initially form within molecular clouds and remain embedded in them. After gas expulsion, the cluster becomes unbound, and gravity from the surrounding residual cloud stretches it into a string-like structure until the cloud disperses.} 
	\label{fig:timeline}
\end{figure}

\begin{figure}
    \centering
    \includegraphics[width=0.8\textwidth]{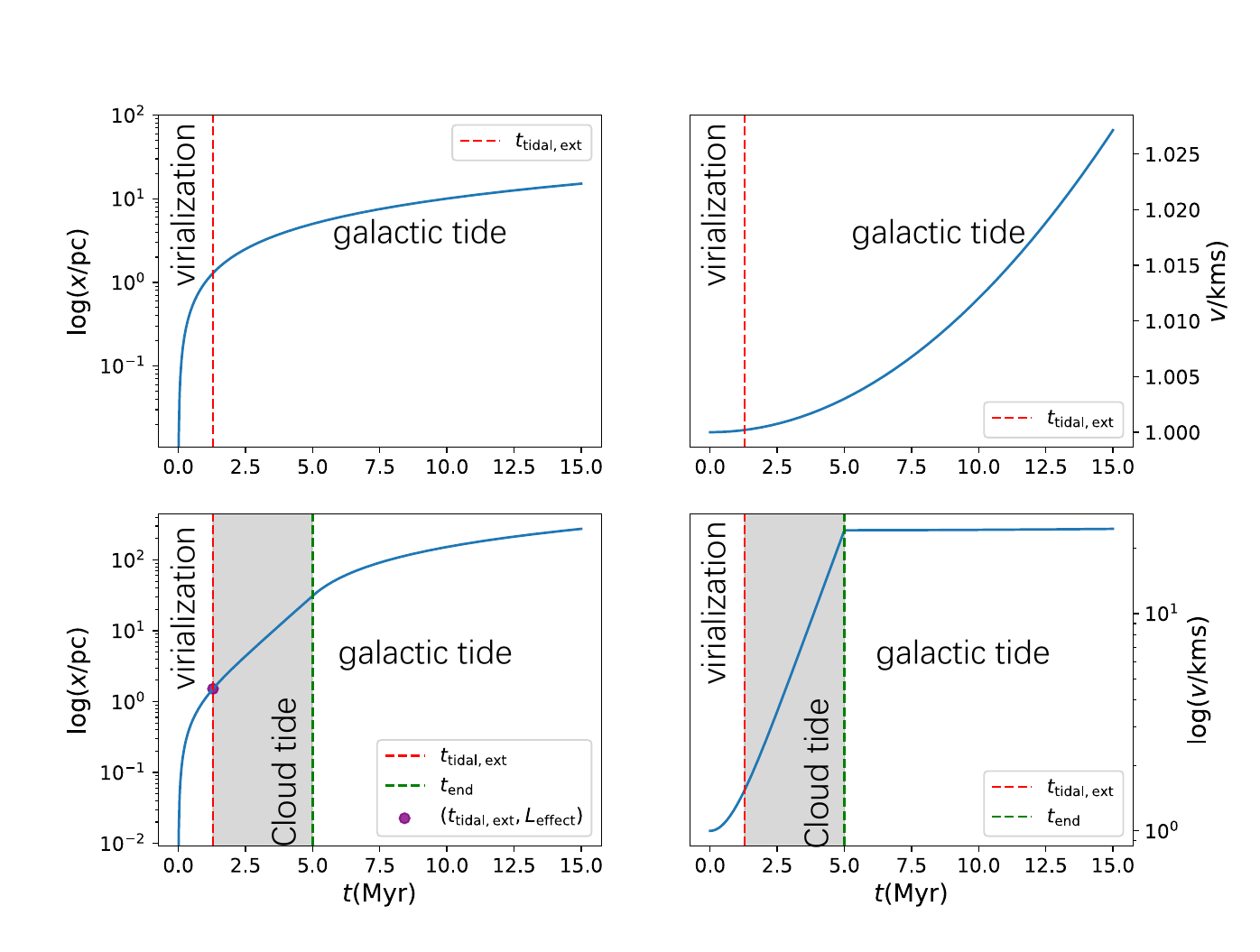}
\caption{Cluster evolution in the absence and presence of cloud tide. The above figure shows the slow evolution without the cloud tide effect, while the bottom figure shows acceleration due to the cloud tide. Cloud tide can lead to significant expansion. In a few Myr, the star cluster can expand from a few parsec to a few tens of parsec.\label{fig:demo}}
\end{figure}

The entire evolutionary process of the cluster should involve a slow expansion due to gas expulsion, followed by tidal-driven expansion caused by remaining gas from the parent cloud, leading to a final integration into the Galaxy. In Fig. \ref{fig:timeline}, we illustrate a timeline of the process.
The effect of the cloud tide is the phase of rapid expansion, which can potentially bring the stellar group from a few parsec to a few tens of parsec, before the parental cloud is completely disrupted by stellar feedback. To analyze this process, we numerically solve the equations controlling radial evolution presented in Sec. \ref{sec:radial} using the initial condition specified in Table. \ref{table}. The results from fiducial simulations, both with and without cloud tide are presented in Fig. \ref{fig:demo}. 


\begin{center}
\begin{tabular}{cccc}
  \toprule
  \textbf{} & \textbf{Parameter} & \textbf{Initial Values} &
  \textbf{Tidal Timescale} \\
  \midrule
  $v_0$ & initial velocity of cluster & $1\;\mathrm{km/s}$ \\
  $x_0$ & initial length of cluster & $0\;\mathrm{pc}$ \\
  $\overline{\rho}_{\mathrm{cloud}}$ & effective density of molecular cloud tide & $10^{-20}\;\mathrm{g/cm^{-3}}$ &  $\sim 0.4\;\mathrm{Myr}$\\
  $\overline{\rho}_{\mathrm{Galactic}}$ & effective density of Galactic tide & $10^{-23}\;\mathrm{g/cm^{-3}}$&  $\sim 14\;\mathrm{Myr}$ \\
  \bottomrule 
  \label{table}
\end{tabular}
\end{center}

To determine the duration through which the cloud tide is effective, we use the results from \citet{zhou2024gascontentevolutionsample}, where they classify different types of molecular clouds; we could calculate the time of stellar feedback using the relative fraction of these objects. Using the equation
\begin{equation}
     t_{\rm `feedback} = t_{\rm YSO}  \times \frac{n_{\rm Type2} + n_{\rm Type3}}{n_{\rm Type1} + n_{\rm Type2} + n_{\rm Type3}}\;,
\end{equation}
We estimate the time of stellar feedback to be 5 Myr. This timescale, although short, can cause the stellar group to reach the size of a few tens of parsecs, which is 10 times larger than the initial size.  We also computed results with different durations where the cloud tide is effective; the results are plotted in Fig. \ref{fig:gaia}.

\subsection{A case of tidally-induced cluster disruption  in Orion molecular cloud}
\begin{figure}
    \centering
    \includegraphics[width=0.90\textwidth]{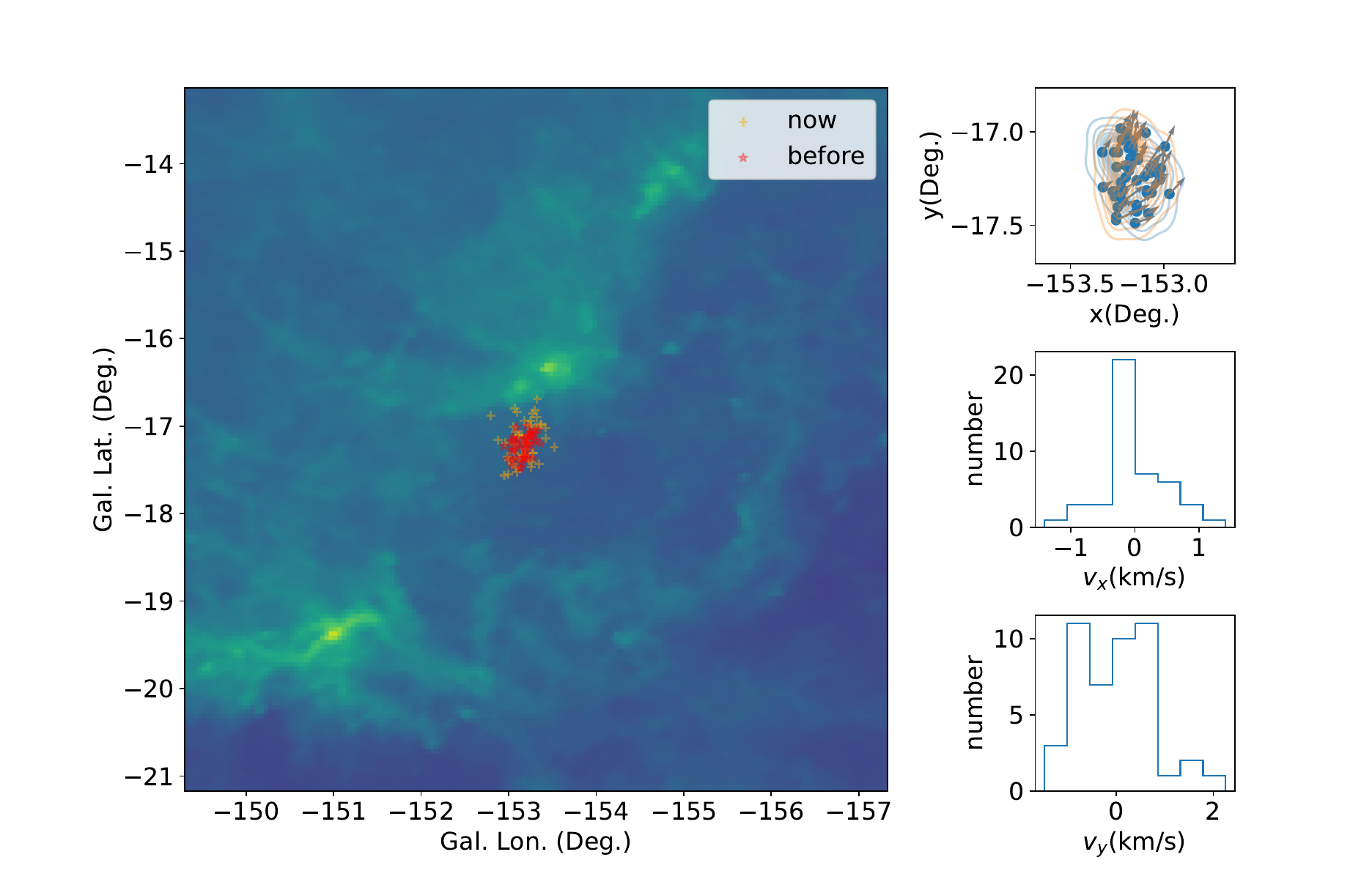}
    \caption{A case of a star cluster under cloud tide found in the Orion molecular cloud. The left Planck plot shows the distribution of gas and the star clusters which is under the influence of the cloud tide. 
    Defining this radial direction as the $y$-axis, we calculated velocity dispersion in all directions (right panel), showing significantly higher dispersion along the longer axis, which points to the nearby Orion B cloud. \label{fig:5}}

\end{figure}

Around the Orion Nebula Cloud (ONC), we identify a cluster whose elongation indicates the importance of cloud tides (Fig. \ref{fig:5}). We used Planck satellite data to map gas distribution and Gaia data to track velocities. For Gaia, we adopted the member star catalog of the Orion structure from \cite{Zhou_2022} which consists of 15149 Class I and Class II YSOs, they selected from \cite{2016MNRAS.458.3479M}. \cite{Zhou_2022} used a machine-learning-based approach so age information is not available and they removed YSOs whose parallax errors are larger than 20 per cent. 
This cluster is a likely case of tidally induced disruption due to its proximity to the Orion B cloud and its elongated shape relative to the cloud.  

Using the cloud direction as the $y$-axis, we analyzed velocities along both axes. The $y$-direction histogram shows significant velocity dispersion compared to the $x$-direction. With the Orion B cloud's mass approximately $10^4 M_{\odot}$ \citep{Pety_2017} and around 3.5 pc the distance adopted,
we assumed an effective cloud tide density of $10^{-19} \, \rm g \, cm^{-3}$. These parameters are similar to what we assuming in Sec. \ref{sec:complete}. This agreement, together with the fact that the elongation of the cluster points to the nearby Orion B cloud, suggests that this can be a case of a star cluster being disrupted by cloud tide. 

\subsection{Relation to Gaia String}
\begin{figure}
    \centering
    \includegraphics[width=0.90\textwidth]{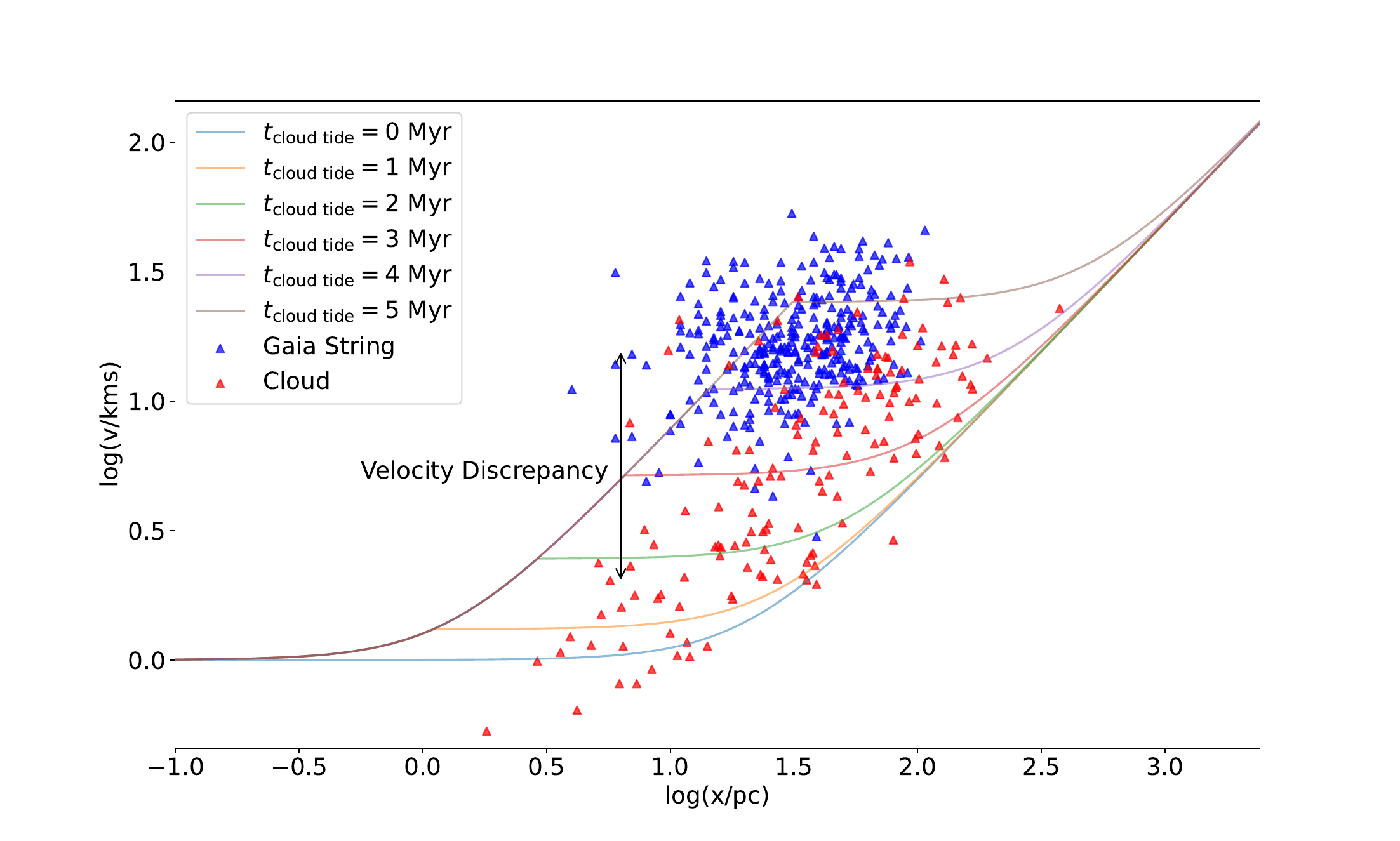}
    \caption{Comparison with other studies. Blue triangles represent clusters
    with string structures from \cite{Zucker_2022}, and red triangles represent
    YSOs from \cite{Zhou_2022}, whose size and velocities tracers that of the
    cloud they are associated with. The Gaia string exhibits a velocity discrepancy
    relative to the molecular cloud, which can be attributed to cloud tide.}
    \label{fig:gaia}
\end{figure}

The cloud tide offers a simple explanation to the existence of a Gaia string, as revealed by recent observations. 
Large, filamentary cloud \citep{Li_2013,Alves_2020,kormann2025supercloudslocalmilkyway,Pfalzner_2016}, when converted into stars, may resemble the Gaia strings. Although the convert of large filamentary structures into star can produce stellar associations that resembles Gaia strings, the cloud tide can also produce elongate stellar associations.
Figure \ref{fig:gaia} compares our cloud tide simulations with string and YSO data. The alignment of string and YSO data near the simulation line suggests that clusters with string shapes experience tidal acceleration from molecular clouds. Gaia strings \citep{2019AJ....158..122K} are stellar associations characterized by elongated, string-like shapes, young ages, and significant lengths. Their morphology indicates a late stage of star cluster evolution, likely resulting from tidally driven dissolution.

To establish the link between cloud tides and Gaia strings, we simulate star cluster disruption by cloud tides, assuming various durations of tidal activation, and compare results with observations in the velocity-dispersion-size plane. Velocities and sizes of YSO associations from \cite{Zhou_2022} are overlaid. From Fig. \ref{fig:gaia}, Gaia strings show larger velocity dispersion than YSO associations, suggesting that mechanisms like cloud tides are necessary. Comparison with our simulations indicates that a cloud tide activation duration of 4 to 5 Myr can explain the properties of these strings, a reasonable timescale. Cloud tides are a plausible mechanism for forming these strings.


\section{Conclusion}

We study the effect of cloud tides on star cluster disruption. Star clusters form in molecular clouds, and after gas expulsion, they are influenced by tides from the parent cloud, which effectively disrupt the cluster as long as it remains close to the cloud.

The tidal strength is described by the tidal tensor
\[
T_{rr} = \frac{8\pi}{3}G \rho_{\rm eff} \;,
\]
where
\[
\rho_{\rm eff} = \frac{m}{\frac{4}{3}\pi r^3}
\]
is the effective gas density at the cluster's location. Calculations show that tidal interactions deform clusters, stretching them along the gradient of the gravitational acceleration. The tidal effects can be summarized as follows:

\begin{itemize}
    \item \textbf{Tidally-driven expansion.} The cluster's evolution is described by
    \[
    \ddot{x} = \frac{8\pi}{3}G \rho_{\rm eff} x \;,
    \]
    with the numerical solution
    \[
    x = \frac{1}{2} \left( x_0 + v_0 t_{\rm tidal,ext} \right) e^{\frac{t}{t_{\rm tidal,ext}}} + \frac{1}{2} \left( x_0 - v_0 t_{\rm tidal,ext} \right) e^{-\frac{t}{t_{\rm tidal,ext}}}\;,
    \]
    which drives accelerated cluster disruption.
    \item \textbf{Tangential oscillation.} The cluster's motion is described by
    \[
    \ddot{x} = -\frac{4\pi}{3}G \rho_{\rm eff} x\;,
    \]
    leading to oscillatory behavior. The numerical solution is
    \[
    x = x_0 \cos\left( \frac{t}{\sqrt{2}t_{\rm tidal,ext}} \right) + \sqrt{2}v_0 t_{\rm tidal,ext} \sin\left( \frac{t}{\sqrt{2}t_{\rm tidal,ext}} \right)\;,
    \]
\end{itemize}
where the tidal timescale is defined as
\[
t_{\rm tidal} = \sqrt{\frac{3}{8\pi G \rho}}, \rho = \rho_{\rm eff}\; \text{or} \; \rho_{\rm mean}\;. 
\]
It is the radial expansion driven by \( T_{rr} \) that can lead to significant expansion of the star cluster.

This mechanism is activated when gas expulsion occurs, and the cluster remains close to the parent cloud.
Under reasonable parameters where \( \rho_{\rm eff} = 10^{-19} \, \rm g \, cm^{-3} \), at a time of a few Myr, cloud tide can transform a star cluster of a few parsecs in size to a few tens of parsecs, leading to sizes similar to that of the Gaia string. Cloud tide is thus an efficient mechanism causing the expansion of the star cluster, and the Gaia string can be a plausible explanation of tidally driven expansion. As cloud tide could elongated the cluster, it particularly relevant for understanding bar-like stellar aggregations.


\section*{Acknowledgment}
We thank our referee for the constructive comments that helped to improve the manuscript.
GXL and XTC acknowledges support from NSFC grant No. 12273032 and
12033005. XTC receives support from the Graduation Innovation Grant No. KC-242410131 from Yunnan University.

\section*{Data Availability}
Table data: The link to the electronic table can be found in \cite{Zhou_2022} and \cite{Zucker_2022}. The full YSO catalogue will be release soon.
Image data: Planck \url{https://pla.esac.esa.int/}.

\bibliography{refer}{}
\bibliographystyle{aasjournal}

\end{document}